# Chiral flow in a binary mixture of two-dimensional active disks

Miguel A. López-Castaño[1]*, A. Rodríguez-Rivas[2] and F. Vega Reyes[1,3]

[1]Departamento de Física, Universidad de Extremadura, Badajoz, Spain, [2]Department of Physical, Chemical and Natural Systems, Pablo de Olavide University, Sevilla, Spain, [3]Instituto de Computación Científica Avanzada (ICCAEx), Universidad de Extremadura, Badajoz, Spain

In this article, we, experimentally, studied the dynamics of a binary mixture of air-fluidized disks. The disks are chiral since they incorporate a set of blades with constant tilt. Both species are identical except for their blade tilt angle, which is rotated by 180° in the second species. We analyzed the phase behavior of the system. Our analysis reveals a wide range of different fluid dynamics, including chiral flow. This chiral flow features in its base state a large vortex. We reported, for certain ranges of relative particle density of each species, inversion of the vorticity of this vortex. We discussed the possible mechanisms behind these chiral flow transitions.

## 1 Introduction

Active particles are defined as macroscopic particles that convert internal or external sources of energy into directed motion [1–6]. Therefore, they configure soft matter systems that are intrinsically out-of-equilibrium. As a consequence, their dynamics gives rise to a wide range of interesting and complex feature-rich collective phenomena; such as motility-induced phase separation [7–9], global chiral flows and vortexes [10–12], rotating crystals [13], and flocking and swarming [4, 14–16]. Furthermore, under certain conditions, active matter may deploy antisymmetric components in the hydrodynamic fluxes. This results in the emergence of a new set of transport coefficients in the constitutive relations [17, 18] as well as odd/Hall diffusion [19–21]. However, the conditions for the emergence of chiral flow and the details of the physical connection between particle chirality and flow chirality are not yet well understood and only very basic description of this issue had been reported so far [22].

However, significant progress has been achieved recently, with the description of the set of steady base chiral flows that an ensemble of identical active particles with constant chirality can display [23]. The newly found variety of chiral flows has been explained theoretically through the emergence of a set of continuous transitions, as recently reported [23]. As it has been shown, these transitions are due to an underlying mechanism of statistical correlations between the rotational and translational degrees of freedom of the active particles. As a consequence, a new fluid flow phase with a chirality with the opposite sign to that of the chirality of the constituent particles has been observed, and we denoted it as the $\mathbb{C}_-$ phase [23]. In another new phase, even more striking, was observed, this consists of a combination of multiple chiral fluid vortexes with vorticities of either sign. We labeled it as *complex chiral phase*, $\mathbb{C}_\pm$. Furthermore, the existence of these statistical correlations has also fundamental effects on transport properties, including diffusion [24].

With respect to this question of chiral flows, but in the context of mixtures of chiral particles (i.e., a set of non-identical chiral particles), there is only a small number of studies, most of them focused on describing phase segregation. For instance, spontaneous phase separation in unbounded assemblies has been reported [13, 16, 25–28]. Moreover, these configurations are prone to display emergent coherent vortexes, ordering and segregation [8, 13, 16]. However, description of flow chirality in active rotors in mixtures is not yet well understood and no chiral flow transitions (in analogy to the monodisperse case [23]) have been, so far, reported. Thus, the motivation of the present work was to detect and characterize analogous transitions that, analogously to the case of a chiral fluid of identical particles, can eventually give rise to a whole new set of chiral flows in a binary mixture of two-dimensional active particles. In particular, we are interested in how the phenomenology of these transitions changes as a function of the relative packing fraction of the species.

For this task, we perform, in this work, a series of experiments with a binary mixture of 3D-printed disks. In our system, the active particles consist of two sets of air-fluidized disks that rotate continuously due to their incorporated tilted blades. All disks are identical except for the blade tilt angle, which differs by 180º between both species. This is so because we are specifically interested in the effects of the predominant particle spin sign (i.e., the *particle chirality* sign). In particular, we will look for chiral flow transitions that can give rise eventually to chiral flow phases analogous to the aforementioned $\mathbb{C}_+$, $\mathbb{C}_\pm$ and $\mathbb{C}_-$ flow phases already observed in the fluid of identical chiral particles.

## 2 Experimental set-up and particle tracking

The set-up consists of a circular air table (diameter $D = 72.5$ cm), delimited by a metallic grid that is perforated with a triangular grid of 3 mm diameter circular holes. The particles were fabricated in our laboratory, by 3D printing. For more details on this 3D printing process, please refer to the Supplementary Material file. Consistent rotation of macroscopic particles can be achieved by means of other experimental methods such as imposing external electromagnetic fields [29, 30] or by means of shape anisotropy in the particles [31].

A constant air current, produced by a fan, comes from below the grid. A polyurethane foam layer at an intermediate height is used to improve the homogeneity. Fan power can be adjusted so that, above a threshold value the upflow speed $u_{air}$, stable levitation of the disks is induced. Upflow past the disks produces turbulent vortexes. This induces Brownian motion on the particles [32] (for more details on the properties of turbulent vortexes, for different obstacle shapes, please refer to [33]). This translational dynamics is contained in this case within the horizontal; thus, it is essentially two-dimensional. In addition, the upflow past the blades also produces a horizontal torque on the particle, the sense of this torque being consistent with the tilt of the blades. For this reason, we have used a tilt angle difference of 180° between both species, which otherwise are almost identical in their other properties (partial diameter $\sigma = 72.5 \pm 0.1$ mm and average mass $m_p = 7.76 \pm 0.01$ g, see Supplementary Material for more details on the mass distribution for each species).

Let us denote the average translational kinetic energy as $\overline{T_t} = (m_p/2)\langle (\mathbf{v} - \mathbf{u})^2 \rangle$, where $\mathbf{u}$ is the velocity flow field (here, $\langle \ \rangle$ refers to the average overall particles and configurations). It is also convenient to define the rotational kinetic energy $\overline{T_r} = (I/2)\langle w^2 \rangle$. In our set-up, the upflow speed $u_{air}$, and $\overline{T_t}, \overline{T_r}$ are coupled since, for each experiment, once the fan power ($u_{air}$) is fixed, a unique steady state with well-defined values of $\overline{T_t}, \overline{T_r}$ is achieved. Thus, for energy-input varying series, we only need to specify the value one these magnitudes. In our case, we will report $\overline{T_t}$.

We characterize, here, the 2D particle density with the packing fraction $\phi = N\sigma^2/D^2$, where $N$ is the total number of disks in the system and $D$ is the diameter of the arena. Henceforth, we will refer to magnitudes for each species with subscripts 1 and 2, respectively. Therefore, we can define the packing fraction $\phi = \phi_1 + \phi_2$, with $\phi_i = N_i\sigma^2/D^2$, $N_i$ being the number of disks of species $i$ present in the experiment. We used the reduced magnitude $\chi = \phi_2 - \phi_1/\phi$, as a measure of the relative particle density of both species 1. For this work, a series of experiments have been carried out, at constant total packing fraction ($\phi = 0.24$), by varying fan power for a set of constant values of relative particle density $\chi$.

We have used a high speed camera to record a total of 160 videos of 100 s, at 900 frames per second and a resolution of $1,280 \times 800$ pixels. Using standard tracking algorithms [34] for the translational movement of the particles and a custom implementation [35] of an iterative cross-correlation technique (applied to an annulus containing the blades of each disk) for the rotational component of the movement, we have been able to accurately keep track of the system dynamics. For a precise description of the methods and uncertainty estimation, please refer to previous works [35].

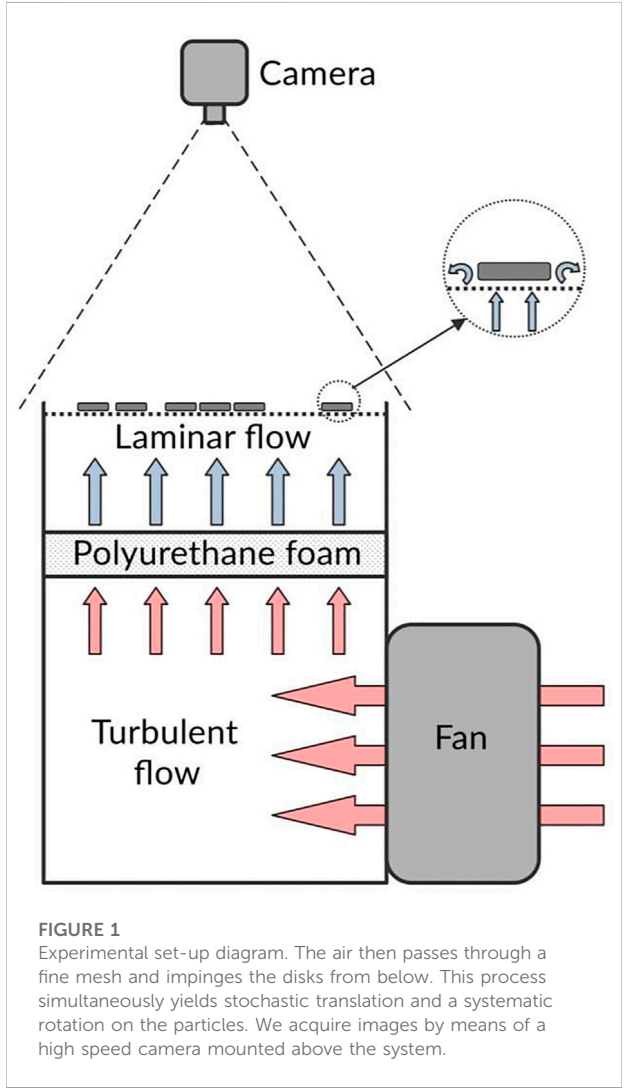

**FIGURE 1**
Experimental set-up diagram. The air then passes through a fine mesh and impinges the disks from below. This process simultaneously yields stochastic translation and a systematic rotation on the particles. We acquire images by means of a high speed camera mounted above the system.

## 3 Results

Figure 2 shows stream lines in steady flows for three different experiments at low $\bar{T}_t$. Similar to the single-component case, we see the emergence of a phase with the same sign of global flow vorticity $\bar{\omega}$ as the average sign of particle spin (determined by $\chi$ as we defined it), which we denote as the $\mathbb{C}_+$ chiral flow phase [23], with one large vortex (left and right figures). Another phase with several smaller vortexes and $\bar{\omega} = 0$ is observed, which we call the complex chiral phase, or $\mathbb{C}_\pm$ (center). In addition, a third phase with different signs of $\bar{\omega}$ and $\chi$ is observed (not represented in that figure), as reported below. This phase is denoted in this work as the $\mathbb{C}_-$ phase. This last phase, as we will see, is obtained for larger energy inputs $\bar{T}_t$.

We also find vorticity sign inversions as average kinetic energy increases. Moreover, we will show that this transition is mediated by an intermediate phase that features complex stable flow vortexes, reminiscent of a turbulent flow. When the system achieves this kind of state, global vorticity approaches zero and partial species vorticities are also neutralized. We find that vortexes are stable and persist along the typical duration of our experiments.

First, we have studied the shape of marginal distribution functions, both for the translational velocity $f_{r,w}(v)$ and for the spin $f_{r,v}(w)$. More specifically, in Figure 3, each color curve represents an experiment at a different average kinetic energy $\bar{T}_t$. We have separated the value of the translational velocity distribution into three columns for studies of spinners of species 1, 2, and combined; as well as in different rows for the each $\chi$ used. We observed a general increase in both the average and variance of the $v$ with increasing $\bar{T}_t$ for all the cases studied, as expected. But a radically different behavior at low $\bar{T}_t$ is observed when we discriminate between disks with different natural spins. A bimodal distribution is observed in the distributions for species 2 at low $\bar{T}_t$ and values close to $\chi = 0$. In contrast, a unimodal distribution is recovered when the system is monodisperse with an increase in velocities close to zero compared to the monodisperse case of species-1 disks.

The effects of particle shape anisotropy and, therefore, natural spin direction, are markedly evident in Figure 4, presenting a bimodal distribution of particle spin velocities. Of particular relevance is the fact that a non-negligible fraction of the angular distribution fraction in some experiments attains values of a sign different to that of the natural spin of the species; by looking at the experimental movies, we have concluded that this "reversal" of the particle spin is caused by particle–particle collisions, especially those collisions in which several particles are involved and some of them remain trapped. The amount of energy dissipation and the non-trivial momentum transfer between rotational and translational degrees of freedom make these collisions very important for explaining the onset of collective chiral flow. As we have stated, the behavior of disks that remain trapped is also reflected in the distribution of rotational speed, with a bimodal distribution in the second species and for a wide range of values, a fact that prevails even for high values of $\bar{T}_t$, recovering a one-modal distribution only in the monodisperse case and for high values of $\bar{T}_t$.

Regarding chirality, active spinners are known to have a strong tendency to display vortex-like collective motion [22, 36]. In our case, we have already studied, in previous works [23] the single-component case and found that, unexpectedly, the chirality of the system experiences a transition from a rotation in the sign of the particle natural spin ($\mathbb{C}_+$ phase) to a counter-spin-wise motion ($\mathbb{C}_-$) without the need for a change in the particle self-rotation sense. This surprising characteristic (surprising in the sense that typically, chiral flows tend to mimic the intrinsic spin of particles), in our current mixture experiments, is present for both species, as it is shown in Figure 5A. Therefore, we wanted to investigate what kind of perturbation would particles of the opposite species produce in the chiral dynamics of our system.

---

1  In this work, species 1 stands for particles rotating clockwise, conversely for species 2. Since the rotation sense is actually not physically relevant, we simply use the numeric subscript notation.

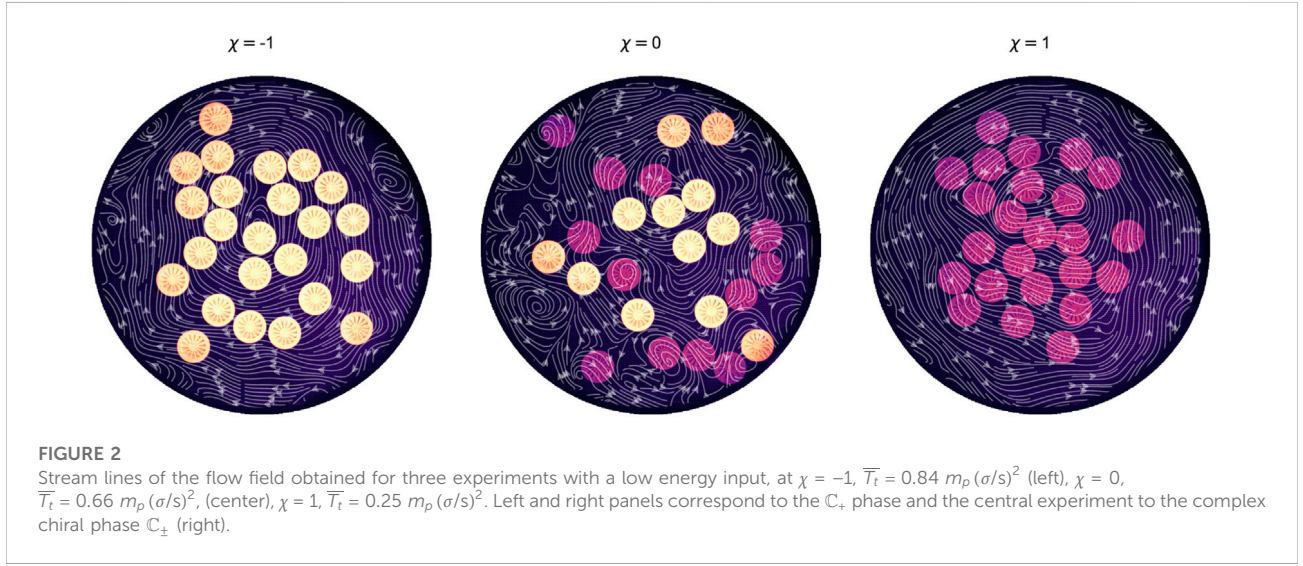

**FIGURE 2**
Stream lines of the flow field obtained for three experiments with a low energy input, at $\chi = -1$, $\overline{T_t} = 0.84\ m_p\,(\sigma/\text{s})^2$ (left), $\chi = 0$, $\overline{T_t} = 0.66\ m_p\,(\sigma/\text{s})^2$, (center), $\chi = 1$, $\overline{T_t} = 0.25\ m_p\,(\sigma/\text{s})^2$. Left and right panels correspond to the $\mathbb{C}_+$ phase and the central experiment to the complex chiral phase $\mathbb{C}_\pm$ (right).

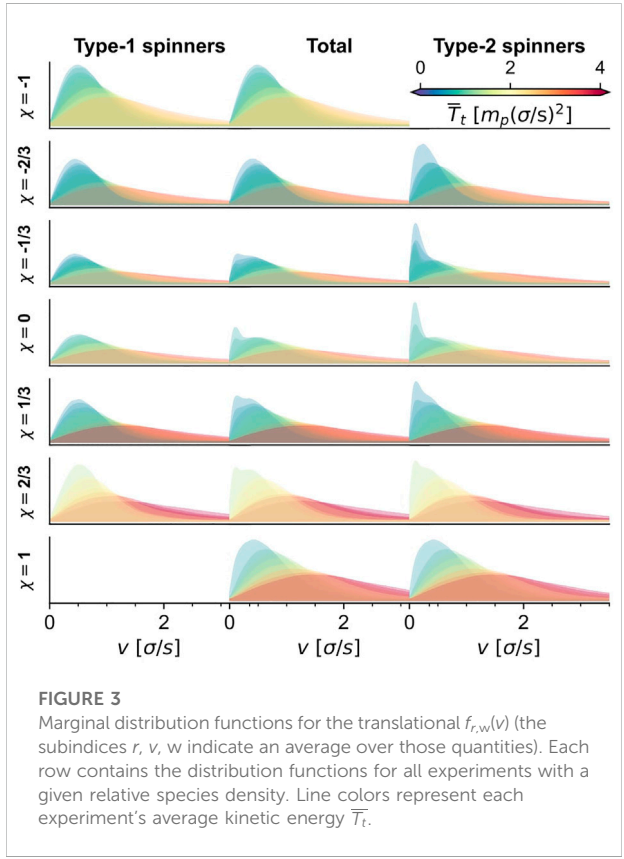

**FIGURE 3**
Marginal distribution functions for the translational $f_{r,w}(v)$ (the subindices $r$, $v$, $w$ indicate an average over those quantities). Each row contains the distribution functions for all experiments with a given relative species density. Line colors represent each experiment's average kinetic energy $\overline{T_t}$.

In order to understand the chirality behavior in our system, we make use of the fluid vorticity, defined here as a field: $\omega = (1/2)\epsilon_{ij}\partial_i u_j$, where $\epsilon_{ij}$ is the 2D Levi-Civita symbol and $u_{i,j}$ are the components of the average fluid velocity field. For each experiment, we have calculated this field and also its average values (both for each species separately and for all particles combined).

We have noticed that there is a clear dependence on the sense and strength of the chiral motion with the average translational kinetic energy. As we mentioned before, in Figure 5A, we show that for monodisperse configurations ($\chi = -1$ or $\chi = +1$); the average vorticity experiences a transition from chirality in the same direction as the particle's natural spin (which we call natural chirality, $\mathbb{C}_+$), for low energy inputs, to displaying a vortex in the opposite direction when the injected energy is high. Also, the trend of fluid vorticity vs translational kinetic energy is opposite for each species, confirming that fluid chirality is closely related to particle dynamical asymmetries (particle chirality).

We must comment, for the sake of transparency, that although we have used the same construction material for both species and checked for mass, width, and length consistency, the dynamical response to the airflow is slightly different for each component with species-2 spinners (naturally CCW rotating) showing a higher mean speed (both translational and rotational) than type-1 spinners for the same value of the input airflow. This is shown in Figure 5B.

We have also observed that the chiral flow is suppressed as the relative fraction of the species approach $\chi \simeq 0$, as shown in Figure 6A, where we represent mean vorticity as a function of the relative density of each species and the average kinetic energy (which is a proxy for spinner activity). Furthermore, we show that the partial vorticities of each species also approach zero when the relative fraction is near zero as Figure 6B shows; this result indicates that the particles of different species do not form separate vortexes as the transition to equal density goes on, they rather experience a gradual decay to a phase with no clear global chirality, probably consisting on a series of several smaller vortexes of different signs ($\mathbb{C}_{\pm}$), see Fig in the Supplementary Material. Thus, the presence of intruder particles breaks the same species particle interactions and correlations that caused chirality in the first place.

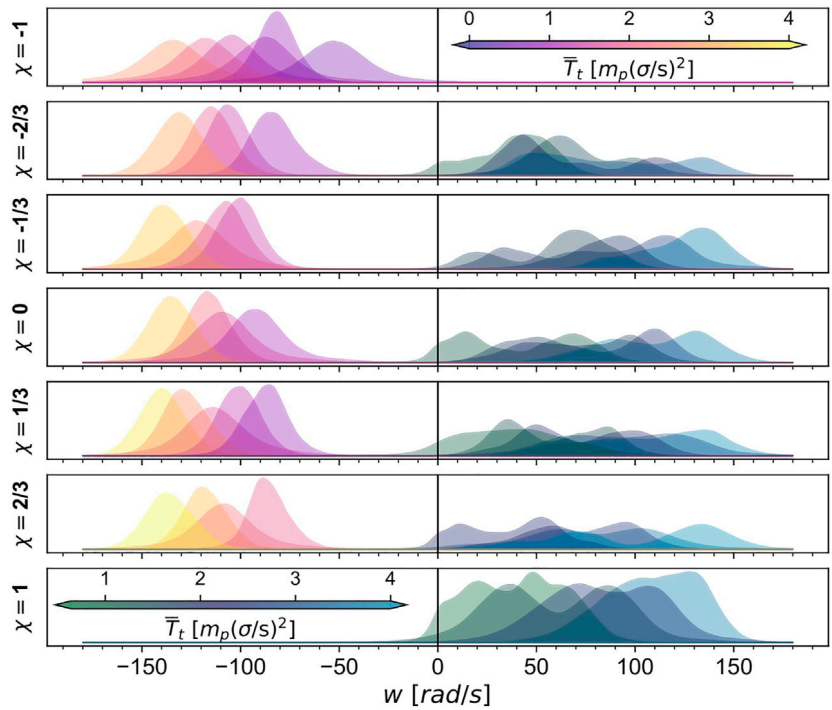

**FIGURE 4**
Marginal distribution function rotational $f_{r,v}(w)$ (right) velocities (the subindices r, v, w indicate an average over those quantities). Each row contains the distribution functions for all experiments with a given relative species density. We show self-spinning distribution functions for the first species spinners (brighter shades of purple represent higher system translational energies) and species-2 spinners (shades of green).

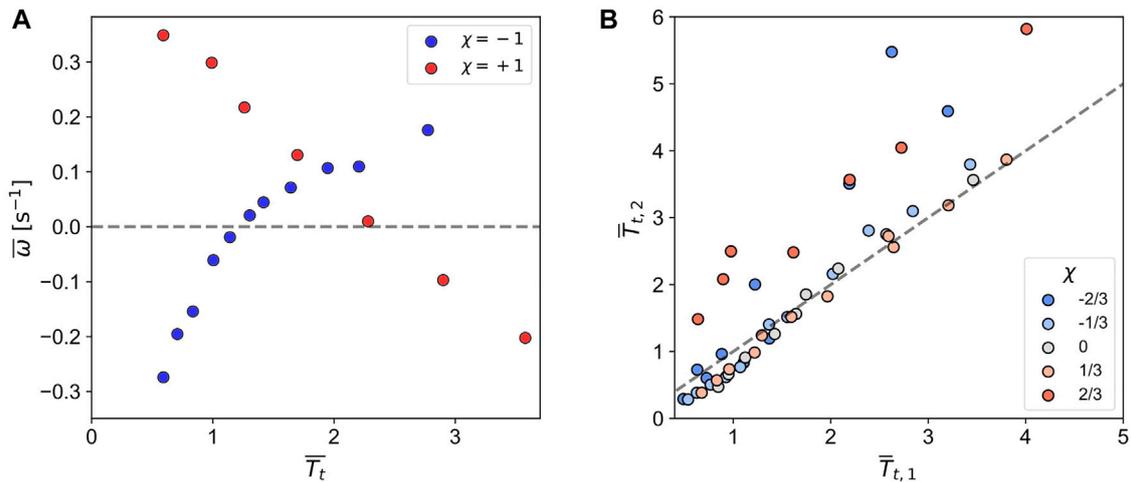

**FIGURE 5**
**(A)** General trend of the chirality for the two single-component cases. **(B)** Mean kinetic energy for both species; type-2 disks tend to show a higher activity level for the same energy input despite our efforts to consistently fabricate both species.

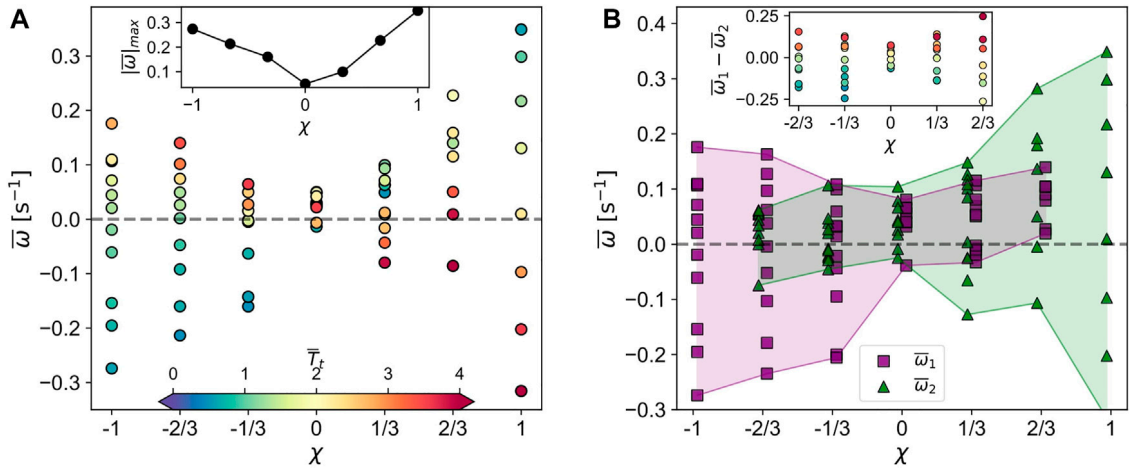

**FIGURE 6**
**(A)** Evolution of the vorticity as a function of $\chi$ and $\overline{T}_t$, absolute values in the inset. **(B)** Partial vorticities of each species, purple for type-1 particles and green for species-2 spinners.

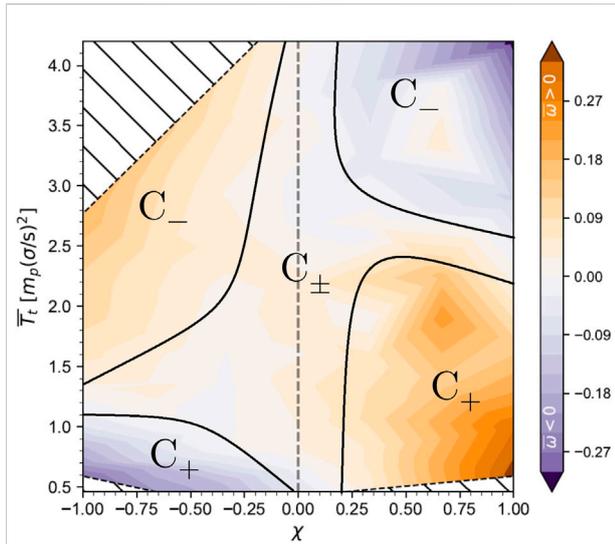

**FIGURE 7**
Average flow vorticity 2D diagram. The sense of chiral flow (sign of $\bar{\omega}$) is indicated by a color gradient, orange means positive vorticity ($\bar{\omega} > 0$), white, $\bar{\omega} = 0$, being the complex phase ($\mathbb{C}_{\pm}$), and purple means negative vorticity. The y-axis represents mean kinetic energy (controlled by the airflow velocity) and x-axis, $\chi$, is the relative density fraction of the species, with $\chi = -1$ being a system composed only of species-1 particles and $\chi = +1$ only kind-2 spinners. We have signaled approximately (solid black lines plus labels) the chiral phase each region represents (phase depends on a combination of the signs $\bar{\omega}$ and $\chi$).

All this phenomenology is condensed in the phase behavior of the system Figure 7 and the difference between species is the root of the slight asymmetry of phases. However, we observe that the results and overall trends we have described here are not affected by this difference between species, showing that the dynamic behaviors studied in [23] are very stable.

Regarding structural ordering, we have represented the well-known radial distribution function in Figure 8A, we can see that the presence of peaks in $g(r)$ is highly correlated with the average translational kinetic energy, which also indicates that states with higher ordering levels correspond to the states where the global chirality mimics the natural spin of particles ($\mathbb{C}_+$ phase). In those cases, the particles tend to be concentrated near the center of the arena and do not explore as much the regions next to the walls, possibly due to a central potential caused by the interaction of the airflow with the boundaries, reminiscent of the force reported in previous works with a similar set-up [32, 37]; this facilitates the creation of more structured ensembles (as shown by the presence of peaks in the $g(r)$ figure). On the other hand, as we increase the energy input, measured by the mean kinetic energy of the particles, the systems evolve to a near gas-like radial distribution function. These trends appear to be independent of the relative density fraction of each species. The behavior of this distribution function $g(r)$ combined with the results presented in Figure 6B indicates that the two observed chiral phases (spin-wise, $\mathbb{C}_+$ and counter-spin-wise, $\mathbb{C}_-$) are fundamentally different in the sense that, when chiral global movement goes in the same direction as particle's natural spin, the system is relatively ordered whereas in the opposite case, particles behave much more like a fluid; further insight is needed in order to determine the role that ordering, and therefore of the collisional frequency, intensity, and correlations, plays on the formation of chirality.

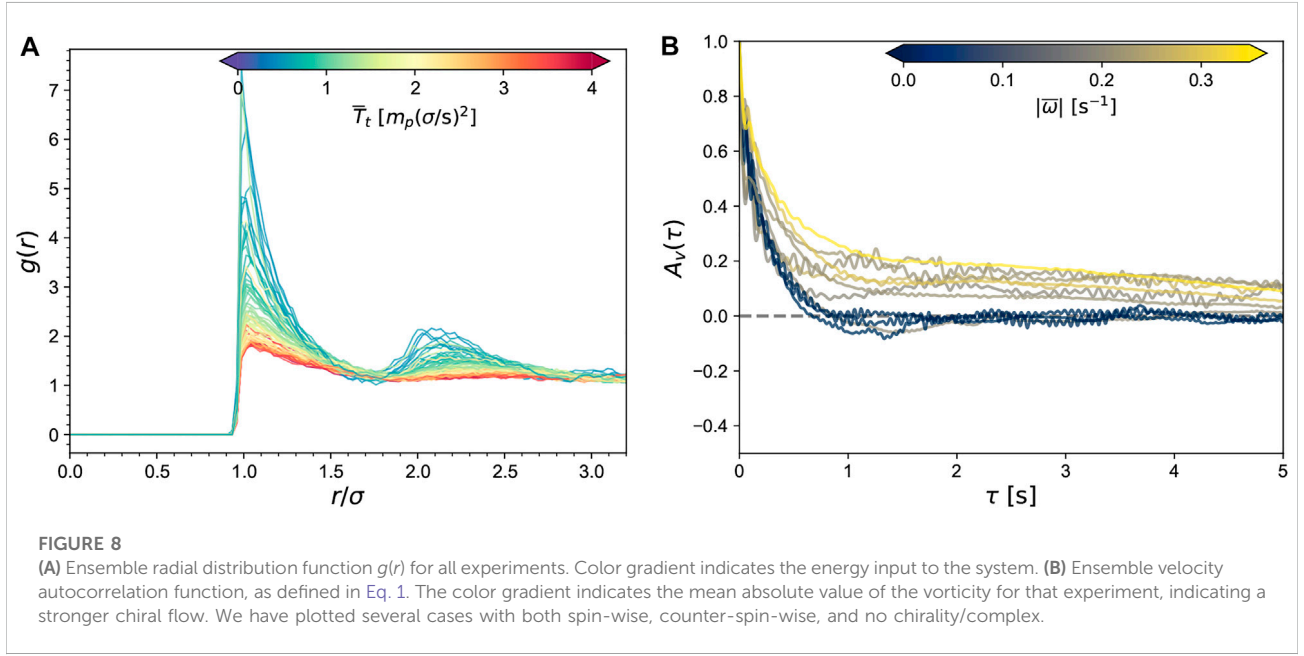

**FIGURE 8**
(A) Ensemble radial distribution function $g(r)$ for all experiments. Color gradient indicates the energy input to the system. (B) Ensemble velocity autocorrelation function, as defined in Eq. 1. The color gradient indicates the mean absolute value of the vorticity for that experiment, indicating a stronger chiral flow. We have plotted several cases with both spin-wise, counter-spin-wise, and no chirality/complex.

A similar interesting behavior is observed in the velocity autocorrelation function. We define this property as

$$A_v(\tau) = \frac{\langle \overline{v_i}(t) \rangle \cdot \langle \overline{v_i}(t + \tau) \rangle}{\langle \overline{v_i}(t) \cdot \overline{v_i}(t) \rangle}, \qquad (1)$$

where $\langle \ \rangle$ correspond to the averaging overall disks $i$ and initial times $t$, with a time step $\tau$. In Figure 8B, we show corresponding correlation curves obtained for different absolute values of the mean value of the vorticity. We have found the relaxation time of this function depends mainly on the absolute value of the average vorticity $\bar{\omega}$ and not so much on the energy input $\overline{T_t}$; that is, when strong chirality is present, velocities remain correlated for longer periods.

In contrast, for a low vorticity field, a rapid decay of the curves (the envelope) is observed, over a high-frequency oscillation. This fact is in good correspondence with the gas and spinners phases of Ref. [31] (Zhang *et. al.*), an article in which a study of the phases and dynamics of a system of rollers is carried out, with small enough particles to consider study zones without the influence of the edges in the experiment. Although they are different types of particles, the global dynamics shows similarities that allow us to compare and show the differences. In the considered study, they relate the frequency of the oscillations with the individual activity of the rollers, which is controlled by an electric field, and which in turn controls the different phases observed and the dynamics of the bearing itself. Similarly, our high-frequency oscillations are characterized by the activity of disks [35]. Compared to the roller system, the slow relaxation curves display lower values for the autocorrelation, mainly due to the action of the boundary conditions together with the implications of considering mixtures of disks of different intrinsic chiralities. Finally, we do not observe a jamming phenomenology. The disks escape from the cages created by their neighbors, with accentuated negative values of auto-correlations.

## 4 Conclusion

As we explained, in a fluid of identical chiral particles, non-trivial transition from a global chiral state that mimics the natural particle spin ($\mathbb{C}_+$ phase) to one rotating counter-spin-wise ($\mathbb{C}_-$) can be observed. We already know that the main driver of this transition is the transfer of moment upon collisions (which transfers information from the particle-level to the system-level) between particles. This is controlled by a change in the statistical correlations between particle spin and translational velocity and, ultimately, we find that the energy input to the set-up influences the sign of these correlations and therefore determines the global chiral state [23].

In the present work, that is for a binary mixture of active disks, each species with opposite particle chirality (here, spin), we show experimentally here that, in general, steady fluid flows display *chirality* very clearly as well. Due to this, the space-averaged vorticity (which we call *global vorticity*, $\bar{\omega}$) is noticeably not null for most of the observed chiral flow states. However, and contrary to what previous bibliography has, in general, considered (see references within the bibliography of a recent review [38]), the sign of $\bar{\omega}$ is not determined by the sign of particle spin. On the contrary, at constant average particle spin (which in our experiments is obtained approximately at constant relative species particle density, $\chi$), we detect a series of transitions in the chiral flow states. They feature inversion of the sign of $\bar{\omega}$ through a continuous transition passing by the intermediate value $\bar{\omega} = 0$, this point being consistently located at $\chi = 0$ (i.e., equal particle density for both species). This means that the fluid composed of a binary mixture of active chiral particles can display all three chiral flow phases already observed for the monodisperse chiral fluid have been observed here again. Interestingly, however, the complex chiral phase occupies, in the binary mixture, significantly larger regions of the parameter space, extending over an area centered in the equimolar mixture case (see $\mathbb{C}_\pm$ regions in Figure 7).

In addition, we have analyzed the structural order and dynamical correlations of the system, by computing the radial distribution function and velocity auto-correlations, respectively. First, it does not appear that species segregation occurs in our system. Also, the radial distribution function reveals that the geometric correlations are completely analogous for binary mixtures with opposite relative particle density $\chi$. Moreover, for the $\mathbb{C}_+$ phase (at low $\overline{T_t}$) the structure of $g(r)$ is more reminiscent to that of a liquid (with secondary maxima). For the $\mathbb{C}_-$ phase, however, $g(r)$ resembles more to the radial distribution function of a gas. On the other hand, the velocity autocorrelation sharply decays in the complex phase $\mathbb{C}_\pm$ while remaining relatively high when strong vortex-like motion is present (high absolute value of the vorticity), independent of the chiral phase.

All of the results combined seem to indicate that the mechanism for chiral flow transition, rather than being determined by boundary conditions, at varying density [22], would be determined by changes in the structure of statistical correlations [23], these changes appear even at constant density. Furthermore, introducing an opposite-spinning species in the system fundamentally is changing the mechanisms that create the correlations giving rise to chiral flow transitions and therefore the phase behavior is even more complex now, as compared to the monodisperse case.

In addition, further work is needed to analyze how these correlations and their rather complex properties are related to the details of momentum and energy transfer upon particle collisions. Thus, it would be interesting to analyze, in future work, the statistical correlations, the vortexes statistics [39] over time, and the flow enstrophy and perform a comparative study of mixtures of active and passive particles.

## Data availability statement

The datasets presented in this study can be found in online repositories. The names of the repository/repositories and accession number(s) can be found at: https://doi.org/10.5281/zenodo.6647705.

## Author contributions

ML-C designed the particles, performed all experiments, wrote most parts of the particle tracking code, and prepared the first version of the manuscript. AR-R co-directed the project and jointly prepared the final version of the manuscript. FVR designed the air table set-up, conceived the experiments, assisted on the implementation of the particle tracking codes, wrote most parts of the post-particle-tracking codes for measurement and characterization of the particle dynamics, edited the final version of the manuscript, and directed and designed the project.

## Acknowledgments

We acknowledge funding from the Government of Spain through Agencia Estatal de Investigación (AEI) project No. PID2020-116567GB-C22). AR-R also acknowledges financial support from Consejería de Transformación Económica, Industria, Conocimiento y Universidades de la Junta de Andalucía through post-doctoral grant no. DC 00316 (PAIDI 2020), co-funded by the EU Fondo Social Europeo (FSE), and from Consejería de Transformación Económica, Industria, Conocimiento y Universidades de la Junta de Andalucía/FEDER through project P20-00816. FVR is also supported by the Junta de Extremadura grant No. GR21091. The authors are indebted to Prof. Ángel Garcimartín and Jeff Urbach, for their important technical and scientific assistance on setting up the laboratory where the experiments of this work were performed.

## Conflict of interest

The authors declare that the research was conducted in the absence of any commercial or financial relationships that could be construed as a potential conflict of interest.

## Publisher's note

All claims expressed in this article are solely those of the authors and do not necessarily represent those of their affiliated organizations, or those of the publisher, the editors, and the reviewers. Any product that may be evaluated in this article, or claim that may be made by its manufacturer, is not guaranteed or endorsed by the publisher.

## Supplementary Material

The Supplementary Material for this article can be found online at: https://www.frontiersin.org/articles/10.3389/fphy.2022.972051/full#supplementary-material

# References

1. Bechinger C, Di Leonardo R, Löwen H, Reichhardt C, Volpe G, Volpe G. Active particles in complex and crowded environments. *Rev Mod Phys* (2016) 88:045006. doi:10.1103/RevModPhys.88.045006

2. Dombrowski C, Cisneros L, Chatkaew S, Goldstein RE, Kessler JO. Self-concentration and large-scale coherence in bacterial dynamics. *Phys Rev Lett* (2004) 93:098103. doi:10.1103/PhysRevLett.93.098103

3. Cavagna A, Cimarelli A, Giardina I, Orlandi A, Parisi G, Procaccini A, et al. New statistical tools for analyzing the structure of animal groups. *Math Biosci* (2008) 214:32–7. doi:10.1016/j.mbs.2008.05.006

4. Zhang HP, Be'er A, Florin EL, Swinney HL. Collective motion and density fluctuations in bacterial colonies. *Proc Natl Acad Sci U S A* (2010) 107:13626–30. doi:10.1073/pnas.1001651107

5. Petroff AP, Wu XL, Libchaber A. Fast-moving bacteria self-organize into active two-dimensional crystals of rotating cells. *Phys Rev Lett* (2015) 114:158102–6. doi:10.1103/PhysRevLett.114.158102

6. Liebchen B, Levis D. *Chiral active matter* (2022).

7. Cates ME, Tailleur J. Motility-induced phase separation. *Annu Rev Condens Matter Phys* (2015) 6:219–44. doi:10.1146/annurev-conmatphys-031214-014710

8. Caporusso CB, Digregorio P, Levis D, Cugliandolo LF, Gonnella G. Motility-Induced microphase and macrophase separation in a two-dimensional active brownian particle system. *Phys Rev Lett* (2020) 125:178004. doi:10.1103/PhysRevLett.125.178004

9. Großmann R, Aranson IS, Peruani F. A particle-field approach bridges phase separation and collective motion in active matter. *Nat Commun* (2020) 11:5365–12. doi:10.1038/s41467-020-18978-5

10. Aref H, B Kadtke J, Zawadzki I, Campbell LJ, Eckhardt B. Point vortex dynamics: Recent results and open problems. *Fluid Dyn Res* (1988) 3:63–74. doi:10.1016/0169-5983(88)90044-5

11. Tsai JC, Ye F, Rodriguez J, Gollub JP, Lubensky TC. A chiral granular gas. *Phys Rev Lett* (2005) 94:214301–4. doi:10.1103/PhysRevLett.94.214301

12. Yang X, Ren C, Cheng K, Zhang HP. Robust boundary flow in chiral active fluid. *Phys Rev E* (2020) 101:022603–6. doi:10.1103/PhysRevE.101.022603

13. Nguyen NH, Klotsa D, Engel M, Glotzer SC. Emergent collective phenomena in a mixture of hard shapes through active rotation. *Phys Rev Lett* (2014) 112: 075701–5. doi:10.1103/PhysRevLett.112.075701

14. Liao GJ, Hall CK, Klapp SH. Dynamical self-assembly of dipolar active Brownian particles in two dimensions. *Soft Matter* (2020) 16:2208–23. doi:10.1039/c9sm01539f

15. Cavagna A, Del Castello L, Dey S, Giardina I, Melillo S, Parisi L, et al. Short-range interactions versus long-range correlations in bird flocks. *Phys Rev E* (2015) 92:012705–15. doi:10.1103/PhysRevE.92.012705

16. Bricard A, Caussin JB, Das D, Savoie C, Chikkadi V, Shitara K, et al. Emergent vortices in populations of colloidal rollers. *Nat Commun* (2015) 6:7470–8. doi:10.1038/ncomms8470

17. Avron JE. Odd viscosity. *J Stat Phys* (1998) 92:543–57. doi:10.1023/a: 1023084404080

18. Banerjee D, Souslov A, Abanov AG, Vitelli V. Odd viscosity in chiral active fluids. *Nat Commun* (2017) 8:1573. doi:10.1038/s41467-017-01378-7

19. Reichhardt C, Reichhardt CJ. Active microrheology, Hall effect, and jamming in chiral fluids. *Phys Rev E* (2019) 100:012604. doi:10.1103/PhysRevE.100.012604

20. Reichhardt C, Reichhardt CJ. Dynamics of Magnus-dominated particle clusters, collisions, pinning, and ratchets. *Phys Rev E* (2020) 101:062602. doi:10.1103/PhysRevE.101.062602

21. Hargus C, Epstein JM, Mandadapu KK. Odd diffusivity of chiral random motion. *Phys Rev Lett* (2021) 127:178001. doi:10.1103/PhysRevLett.127.178001

22. Workamp M, Ramírez G, Daniels KE, Dijksman JA. Symmetry-reversals in chiral active matter. *Soft Matter* (2018) 14:5572–80. doi:10.1039/c8sm00402a

23. López-Castaño MA, Seco AM, Seco AM, Rodríguez-Rivas Á, Reyes FV. *Chirality transitions in a system of active flat spinners* (2021). Preprint in arXiv: 2105.02850.

24. Lopez-Castaño MA, Rodriguez-Rivas A, Vega Reyes F. *Diffusive regimes in a two-dimensional chiral fluid* (2022). Preprint on arXiv:2202.08920.

25. Maloney RC, Liao GJ, Klapp SH, Hall CK. Clustering and phase separation in mixtures of dipolar and active particles. *Soft Matter* (2020) 16:3779–91. doi:10.1039/c9sm02311a

26. Rogel Rodriguez D, Alarcon F, Martinez R, Ramírez J, Valeriani C. Phase behaviour and dynamical features of a two-dimensional binary mixture of active/passive spherical particles. *Soft Matter* (2020) 16:1162–9. doi:10.1039/c9sm01803d

27. Van Der Meer B, Prymidis V, Dijkstra M, Filion L. Predicting the phase behavior of mixtures of active spherical particles. *J Chem Phys* (2020) 152:144901. doi:10.1063/5.0002279

28. Buttinoni I, Bialké J, Kümmel F, Löwen H, Bechinger C, Speck T. Dynamical clustering and phase separation in suspensions of self-propelled colloidal particles. *Phys Rev Lett* (2013) 110:238301–5. doi:10.1103/PhysRevLett.110.238301

29. Grzybowski BA, Stone HA, Whitesides GM. Dynamic self-assembly of magnetized, millimetre-sized objects rotating at a liquid-air interface. *Nature* (2000) 405:1033–6. doi:10.1038/35016528

30. Tierno P, Snezhko A. Transport and assembly of magnetic surface rotors**. *ChemNanoMat* (2021) 7:881–93. doi:10.1002/cnma.202100139

31. Zhang B, Sokolov A, Snezhko A. Reconfigurable emergent patterns in active chiral fluids. *Nat Commun* (2020) 11:4401. doi:10.1038/s41467-020-18209-x

32. Ojha RP, Lemieux PA, Dixon PK, Liu AJ, Durian DJ. Statistical mechanics of a gas-fluidized particle. *Nature* (2004) 427:521–3. doi:10.1038/nature02294

33. [Dataset] van Dyke M, White FM. An album of fluid motion. *J Fluids Eng Dec* (1982) 104(4):542–3. doi:10.1115/1.3241909

34. Crocker JC, Grier DG. Methods of digital video microscopy for colloidal studies. *J Colloid Interf Sci* (1996) 179:298–310. doi:10.1006/jcis.1996.0217

35. López-Castaño MÁ, Márquez-Seco A, Márquez-Seco A, Rodríguez-Rivas ÁVega Reyes F. Fast measurement of angular velocity in air-driven flat rotors with periodical features. *arxiv* (2021).

36. Farhadi S, Machaca S, Aird J, Torres Maldonado BO, Davis S, Arratia PE, et al. Dynamics and thermodynamics of air-driven active spinners. *Soft Matter* (2018) 14: 5588–94. doi:10.1039/c8sm00403j

37. Ojha RP, Abate AR, Durian DJ. Statistical characterization of the forces on spheres in an uflow of air. *Phys Rev E* (2005) 71:016313–7. doi:10.1103/PhysRevE.71.016313

38. Bowick MJ, Fakhri N, Marchetti MC, Ramaswamy S. Symmetry, thermodynamics, and topology in active matter. *Phys Rev X* (2022) 12:010501. doi:10.1103/PhysRevX.12.010501

39. Seo SW, Ko B, Kim JH, Shin Y. Observation of vortex-Antivortex pairing in decaying 2D turbulence of a superfluid gas. *Sci Rep* (2017) 7:4587–8. doi:10.1038/s41598-017-04122-9

40. Lopez-Castaño MA, Rodriguez-Rivas A, Vega Reyes F. Chiral flow in a binary mixture of two-dimensional active disks - supplementary Data. *Tech Rep* (2022). doi:10.5281/zenodo.6647705